\documentclass{article}
\usepackage{amsmath,amssymb,graphicx}
\usepackage{natbib}

\newenvironment{affiliations}{%
  \newcommand\aff[1]{%
    \par\noindent$^{##1}$%
  }\it}{\normalfont}

\title{Insight into the microphysics of antigorite deformation from spherical nanoindentation}
\author{Lars N. Hansen$^1$, Emmanuel C. David$^2$, Nicolas Brantut$^2$, David Wallis$^3$}
\date{\ }

\begin{document}

\maketitle

\begin{affiliations}
\aff{1} Department of Earth Sciences, University of Oxford, Oxford, UK.
\aff{2} Department of Earth Sciences, University College London, London, UK.
\aff{3} Department of Earth Sciences, Utrecht University, Utrecht, The Netherlands.
\end{affiliations}

\begin{abstract}
  The mechanical behavior of antigorite strongly influences the strength and deformation of the subduction interface. Although there is microstructural evidence elucidating the nature of brittle deformation at low pressures, there is often conflicting evidence regarding the potential for plastic deformation in the ductile regime at higher pressures. Here, we present a series of spherical nanoindentation experiments on aggregates of natural antigorite. These experiments effectively investigate the single-crystal mechanical behavior because the volume of deformed material is significantly smaller than the grain size. Individual indents reveal elastic loading followed by yield and strain hardening. The magnitude of the yield stress is a function of crystal orientation, with lower values associated with indents parallel to the basal plane. Unloading paths reveal more strain recovery than expected for purely elastic unloading. The magnitude of inelastic strain recovery is highest for indents parallel to the basal plane. We also imposed indents with cyclical loading paths, and observed strain energy dissipation during unloading-loading cycles conducted up to a fixed maximum indentation load and depth. The magnitude of this dissipated strain energy was highest for indents parallel to the basal plane. Subsequent scanning electron microscopy revealed surface impressions accommodated by shear cracks and a general lack of lattice misorientation around indents, indicating the absence of dislocations. Based on these observations, we suggest that antigorite deformation at high pressures is dominated by sliding on shear cracks. We develop a microphysical model that is able to quantitatively explain the Young's modulus and dissipated strain energy data during cyclic loading experiments, based on either frictional or cohesive sliding of an array of cracks contained in the basal plane.
\end{abstract}

\section{Introduction}

Antigorite is one of the dominant hydrous phases in oceanic lithosphere associated with subduction zones, and its mechanical behaviour plays a key role in controlling the strength of the subduction interface (e.g., Reynard 2013). Experimental investigations of the rheology of antigorite have revealed a number of unique characteristics. First, in the brittle regime at confining pressures less than 400 MPa and at room temperature, antigorite aggregates experience little to no pre-failure dilatancy or stress-induced anisotropy in seismic-wave velocities \citep{escartin97,david18}, in sharp contrast to other low-porosity crystalline rocks \citep{paterson05}. Second, the ductile regime is apparently limited to a high pressure, low temperature domain, with a transition back to brittle, unstable behaviour as temperature increases towards the dehydration temperature of antigorite \citep{chernak10,proctor16}. Under typical subduction zone conditions, at confining pressures of several gigapascals and temperatures of $\sim400^\circ$C, experimental observations are inconclusive regarding the dominant rheological behavior. \citet{hilairet07} and \citet{amiguet12} report power-law creep behaviour consistent with dislocation creep, whereas the results from \citet{proctor16} indicate very high stress exponents that are more consistent with exponential creep and plasticity. In all experiments conducted under elevated pressures and temperatures, the tendency towards strain localisation appears to complicate the interpretation of macroscopic stress-strain behavior. According to experimental data from \citet{chernak10} and \citet{auzende15}, it remains unclear whether it is even possible for antigorite to deform in a fully crystal-plastic regime.

Some insight can be gained into the microphysics of antigorite deformation from the microstructures produced during deformation. Field observations of deformed, antigorite-rich rocks exhumed from subduction-zone environments typically reveal strong foliations \citep[e.g.,][]{hermann00,padron-navarta12}. Foliated antigorite also often exhibits a strong crystallographic preferred orientation (CPO) with (001) mostly parallel to the foliation, which has been interpreted as a marker of flow by dislocation creep, although specific slip systems are still debated \citep{padron-navarta12}. Microstructural observations of both experimentally and naturally deformed antigorite tend to indicate that the crystallographic structure of antigorite, with a corrugation of the (001) plane in the [100] direction, might prevent dislocation glide in the basal plane \citep{auzende15}. In addition, recent deformation experiments conducted on antigorite single crystals with in situ electron microscopy demonstrate that cleavage opening, delamination, and fracture might be the dominant intracrystalline deformation processes \citep{cordier18}. These experimental data are not necessarily in contradiction with the observation of strong CPO in naturally deformed antigorite, considering that a CPO might originate from cleavage along basal planes and associated grain rotation. Overall, the deformation mechanisms of antigorite remain unconstrained, and only indirect evidence for the operation of dislocation creep has been obtained. Cleavage and delamination along the basal plane has been widely reported in experimentally deformed samples, but it remains unclear whether dislocation activity could become dominant under geological strain rates.

To gain further insight into the intragranular deformation mechanisms of antigorite, we conducted a series of nanoindentation experiments. This deformation technique spontaneously generates confining pressure and has been used to study low-temperature, crystal-plastic deformation mechanisms in rock-forming minerals \citep{evans79,basu09,kumamoto17}. Because of the extremely small scale of these deformation experiments, this technique is well suited to investigate the inelastic deformation of antigorite single crystals. We investigate elastic loading, yield, and static internal friction as a function of crystallographic orientation and then discuss the potential microphysical processes operating during deformation of antigorite.

\section{Methods}
\subsection{Sample material and preparation}

Mechanical characterization was carried out on a natural antigorite serpentinite. Serpentinite blocks were acquired from the Rochester quarry of Vermont Verde Antique. Our material is sourced from a block from which similar material was characterized by \citet{david18}. Material from a similar origin and location has been characterized in previous work \citep{reinen94, escartin97a, escartin97, chernak10}. This serpentinite is primarily composed of antigorite ($>95$\%) with minor amounts of magnetite and magnesite. We worked specifically on a 1~cm$\times$1~cm$\times$0.3~cm section cut from a larger core sample that was originally investigated by \citep{david18,david19}. The section was cut normal to the antigorite foliation.

The section was ground and polished to yield a surface that was as flat and smooth as possible. The sample section was first bonded onto an aluminum cylinder using a thermoplastic cement (Crystalbond\texttrademark 509). Initial grinding was conducted with a bonded diamond grinding wheel. Subsequent polishing was conducted on lapping clothes with diamond suspensions of progressively finer grit size, down to a grit size of 0.05~$\mu$m.

\subsection{Spherical nanoindentation}
\subsubsection{Experimental protocol}

Nanoindentation was carried out using an MTS NanoIndenter XP equipped with continuous stiffness measurement (CSM). Indentation tests were performed with a conospherical diamond tip with a nominal tip radius of 10~$\mu$m. Two sets of indents were created on two different areas of the sample section, each with a different methodology.

In the first area, we imposed an array of 8$\times$6 indents spaced on a 50-$\mu$m grid. An initial series of indentation was performed to 100~nm depth, immediately followed by an additional series in the same locations as the previous indents but to 500~nm depth. The initial 100-nm indents were performed to allow easier estimation of the sample modulus, as detailed below, in the same locations as inelastic deformation was induced during the 500-nm indents. Indentation was controlled at a constant indentation strain rate of 0.05~s$^{–1}$, where strain rate is defined as the loading rate divided by the load. Once the maximum depth was reached, the indenter was immediately unloaded at the same rate until a load of 1.5~mN was reached. At this point, the indenter load was held constant and the indenter position was monitored to assess any thermal drift associated with temperature changes inside the indenter housing. Throughout each indent, we recorded the indenter load, displacement, and contact stiffness (via CSM), although contact stiffness was not continuously measured during unloading.

In the second area, we imposed two arrays of 9$\times$5 indents spaced on a 50-$\mu$m grid. For this data set, six loading cycles were performed at each grid point. On the first cycle, the load was increased until a load of 5~mN was reached, at which point the indenter was unloaded to 1.5~mN. The loading was then repeated five more times following the same procedure but progressively increasing the maximum load to 9, 19, 38, 75, and 150~mN on each successive cycle. Loading and unloading were controlled at constant rate of 1.5~mN/s, and a hold was performed on the final unload to assess thermal drift as described above. Throughout each set of cycles, we recorded the indenter load and displacement but did not continuously record the contact stiffness.

\subsubsection{Analysis of indentation data}
\label{sec:process}

Spherical nanoindentation has been a popular characterization technique for a wide variety of materials. As opposed to indentation with sharp tips (e.g., Berkovich), spherical indentation benefits from (1) an initial contact that is purely elastic, (2) analytical solutions for the stress and strain distributions during elastic loading, and (3) an easily identifiable transition between elastic and plastic deformation \citep[e.g.,][]{basu06,field93,angker06}.

\begin{figure}
  \centering
  \includegraphics[scale=0.6]{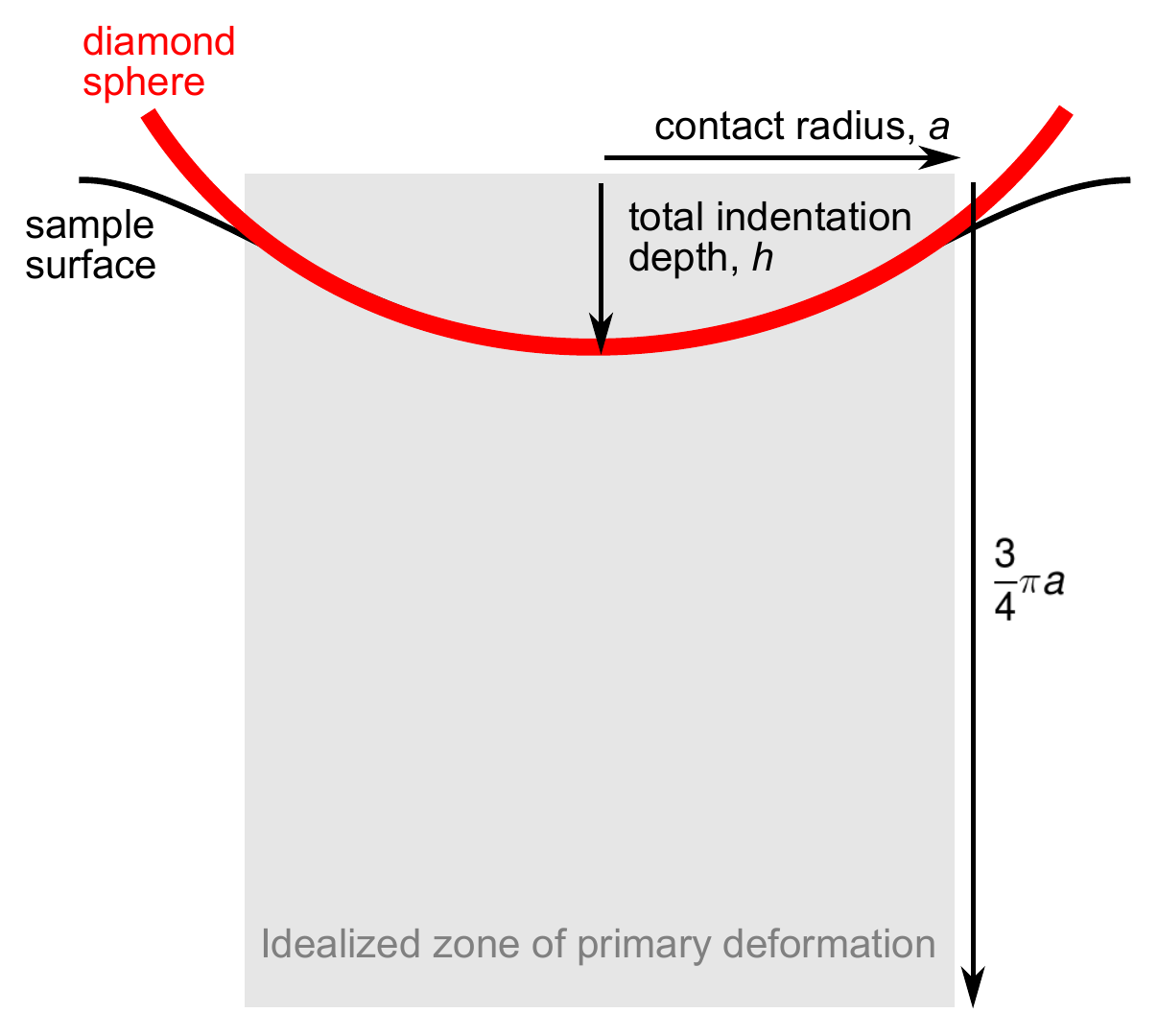}
  \caption{%
    Schematic diagram of spherical indentation. The sample surface (black) is deflected by a spherical indenter tip (red). The gray region depicts the cylindrical region assumed to be the primary region undergoing deformation and used to calculate stress and strain.}
  \label{fig:1}
\end{figure}

The basic configuration of spherical indentation is described in Figure \ref{fig:1}. The mechanics of a spherical contact were originally derived by \citet{hertz82} and are typically presented as
\begin{equation}\label{eq:P}
  P = \frac{4}{3}E_\mathrm{eff}R_\mathrm{eff}^{1/2}h_\mathrm{e},
\end{equation}
where $P$ is the load on the contact, $E_\mathrm{eff}$ is the effective modulus of the contact, $R_\mathrm{eff}$ is the effective radius of curvature of the indenter, and $h_\mathrm{e}$ is the elastic portion of the indentation depth. Assuming the contact is totally elastic, $h_\mathrm{e}$ is equal to the total indentation depth, $h_\mathrm{t}$. The projected area of contact (Figure \ref{fig:1}) is defined by the contact radius, 
\begin{equation} \label{eq:a}
  a = \sqrt{R_\mathrm{eff}h_\mathrm{e}},
\end{equation}
and the contact stiffness, $S = dP/dh_\mathrm{e}$, is therefore
\begin{equation}\label{eq:S}
  S = 2aE_\mathrm{eff}.
\end{equation}

In general, $h_\mathrm{e}$, is only directly measurable if the indentation is known (or assumed) to be entirely elastic, which is typically the case for unloading segments. CSM works by superimposing a small, high-frequency oscillation on top of the primary loading, which effectively consists of many elastic unloading segments and allows $S = dP/dh_\mathrm{e}$ to be continually measured.

Much effort has been put into using the above relationships to produce stress-strain curves from indentation data \citep{herbert01, bushby01,pathak15}. We follow the method reviewed by \citet{pathak15}, in which the indentation stress is defined as the load over the contact area,
\begin{equation} \label{eq:sigma}
  \sigma = \frac{P}{\pi a^2},
\end{equation}
and the indentation strain is defined as compression of an idealized, cylindrical zone of radius $a$ and height $3\pi a/4$ (Figure \ref{fig:1}),
\begin{equation}\label{eq:epsilon}
  \epsilon = \frac{4}{3\pi}\frac{h_\mathrm{e}}{a}.
\end{equation}
These definitions are designed to ensure that the initial elastic segments of the resultant stress-strain curves are in agreement with the elastic modulus of the sample.

In our single indents in the first area, we record data for $P$, $h_\mathrm{t}$, and during loading, $S$ via the CSM. Key unknowns are therefore $R_\mathrm{eff}$ and $E_\mathrm{eff}$. The radius of the indenter tip, $R_\mathrm{i}$, is determined through calibration indents on elastic standards with known modulus (e.g., fused silica) and is related to the effective radius by $R_\mathrm{eff}^{-1} = R_\mathrm{i}^{-1} + R_\mathrm{s}^{-1}$, where $R_\mathrm{s}$ is the radius of curvature of the sample surface (normally taken to be infinity). With known $R_\mathrm{eff}$, we then find the value of $E_\mathrm{eff}$ that best fits a segment of our data shortly after contact that we assume is fully elastic. $E_\mathrm{eff}$ is related to the sample modulus by $E_\mathrm{eff}^{-1} = (1-\nu_\mathrm{i}^2)/E_\mathrm{i} + (1-\nu_\mathrm{s}^2)/E_\mathrm{s}$  , where $E_\mathrm{i}$ is the elastic modulus of the diamond indenter, and $E_\mathrm{s}$ is the elastic modulus of the sample. $E_\mathrm{s}$ is essentially the Young's modulus of the sample in the direction of loading, although the elastic anisotropy tends to be underestimated in highly anisotropic materials \citep[e.g.,][]{kumamoto17}. With $R_\mathrm{eff}$ and $E_\mathrm{eff}$ known, equations \eqref{eq:S}, \eqref{eq:sigma}, and \eqref{eq:epsilon} can be used to generate stress-strain curves.

As will be described below, the unloading portions of stress-strain curves are useful in interpreting our results on antigorite. We did not collect CSM data during unloading (as per standard operating methods), which means we do not have continuous measurements of 
$S$ to estimate a during unloading. However, a can still be estimated from equation \eqref{eq:a} as long as $R_\mathrm{eff}$ is known. Unfortunately, if inelastic deformation has occurred, $R_\mathrm{s}$ will no longer be infinite due to the residual impression that has formed. Therefore, we estimate $R_\mathrm{eff}$ at the end of loading by fitting equation \eqref{eq:P} to an initial segment of the unloading data and assuming the sample modulus is unchanged. We then use this new value of $R_\mathrm{eff}$ and equation \eqref{eq:a} to estimate the contact radius, stress, and strain during unloading.

For cyclical indentation experiments in the second area, we do not have any CSM data and therefore cannot estimate the contact area, stress, or strain for most of the deformation path. However, we can measure $S$ at the beginning of each unloading cycle and obtain an estimate of $a$ through equation \eqref{eq:S}, and therefore get a single measurement of stress and strain for each cycle. We estimate the elastic modulus of the sample in the same manner as described above. In addition, we are interested in the energy budget during each loading cycle. The total energy input into the system is simply the integral of the load-displacement curve. We specifically look for differences between the energy recovered during unloading and the energy input on subsequent reloading to the same load and depth. We calculate the energy difference for this portion of each loading cycle and normalize that value by the volume of the idealized cylinder of deformation (Figure \ref{fig:1}) represented by the contact radius measured during the initial unloading.

\subsection{Microstructural characterization}

Subsequent to indentation, we characterized the microstructure in the vicinity of indentation arrays using scanning electron microscopy (SEM) and electron backscatter diffraction (EBSD). Regions of interest were mapped with either an FEI Quanta 650 FEG E-SEM housed at the University of Oxford or with a Philips XL-30 housed at the University of Utrecht, both equipped with an Oxford Instruments Nordlys-Nano EBSD camera and AZtec 3.3 acquisition software. EBSD data were acquired with an accelerating voltage of 30 kV and currents on the order of 10 nA. Diffraction patterns were obtained with either 2x2 or 4x4 binning of pixels of the EBSD detector. Diffraction patterns were indexed by comparison to a match unit based on a crystal structure modified from \citet{capitani06}. The $\alpha$ angle is very sensitive to composition and was therefore adjusted by modifying the length of the a lattice parameter to 80 Å based on preliminary tests to achieve optimal indexing.The electron beam was rastered across the sample using step sizes of 0.2--0.4~$\mu$m.

\begin{figure}
  \centering
  \includegraphics[width=\textwidth]{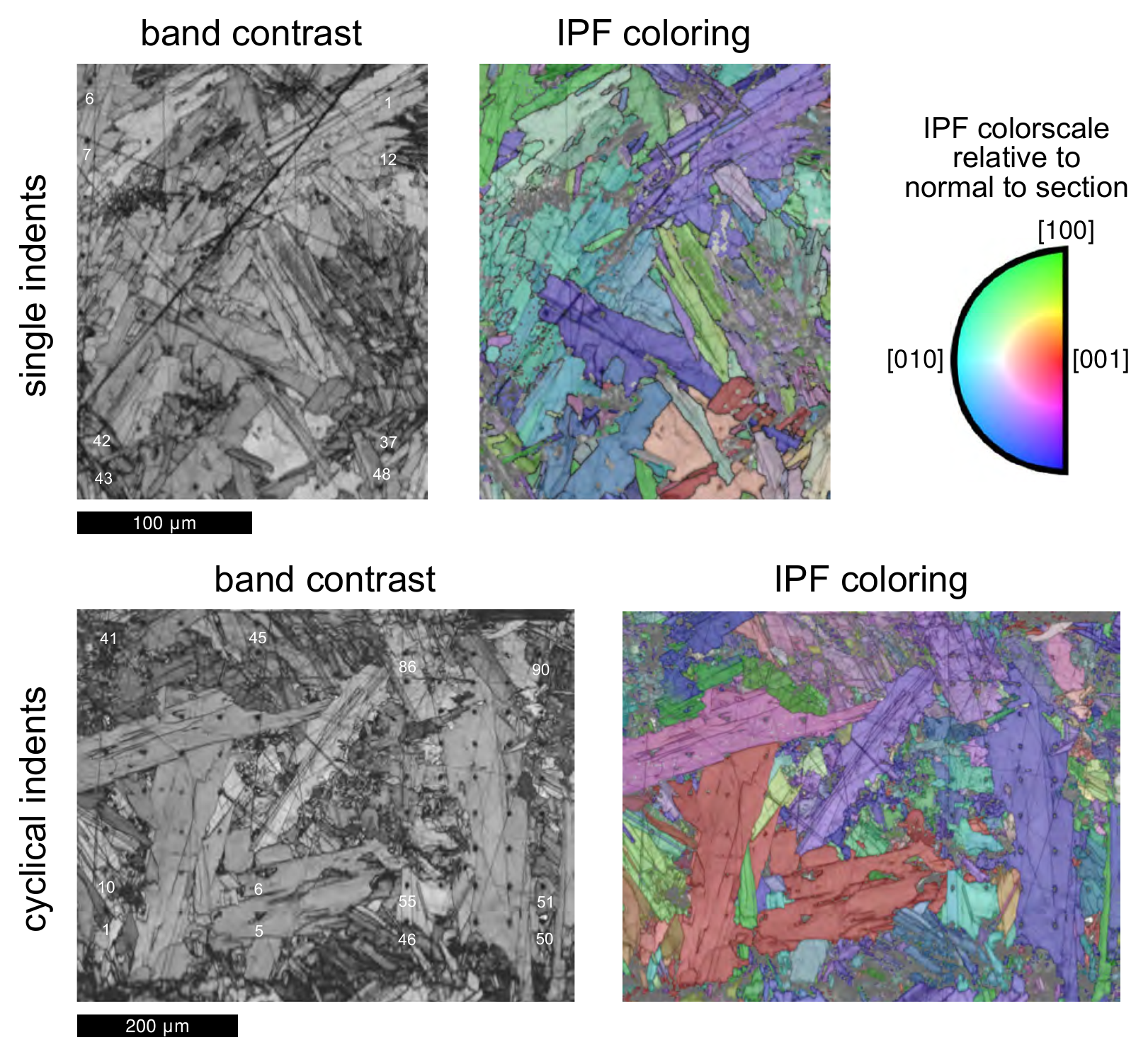}
  \caption{%
    EBSD maps of indent locations. The top row presents maps of the first area in which single indents were placed. The bottom row presents maps of the second area in which cyclical indents were placed. Band contrast maps are presented with white labels to indicate the indent numbering scheme. Additional maps are presented colored according to an inverse pole figure (IPF) for the direction normal to the sample surface (i.e., the direction of indentation). Colored maps are transparently overlain on top of band contrast maps. Black lines indicate calculated grain boundaries.}
  \label{fig:2}
\end{figure}

To index antigorite, we used custom-built match units derived from the diffraction analyses of \citet{capitani06}. Rates of successful indexing were approximately 75\%. In post-processing, isolated individual pixels with no orientation relationship to surrounding pixels were removed. Pixels that were not indexed were filled with the average orientation of neighboring pixels if they had 5 or more indexed nearest neighbors. The resulting EBSD maps are presented in Figure \ref{fig:2}, with the numbering of individual indents annotated for reference.

Individual indents were also imaged with SEM. We collected secondary-electron images with the sample inclined 70$^\circ$ relative to the electron beam, as is typically used for EBSD mapping. Images are presented with a tilt correction to account for foreshortening. Electron imaging in this configuration increases shadowing around topographic features to emphasize surface characteristics.
  
\section{Results}
\subsection{Single indents}

\begin{figure}
  \centering
  \includegraphics[width=\textwidth]{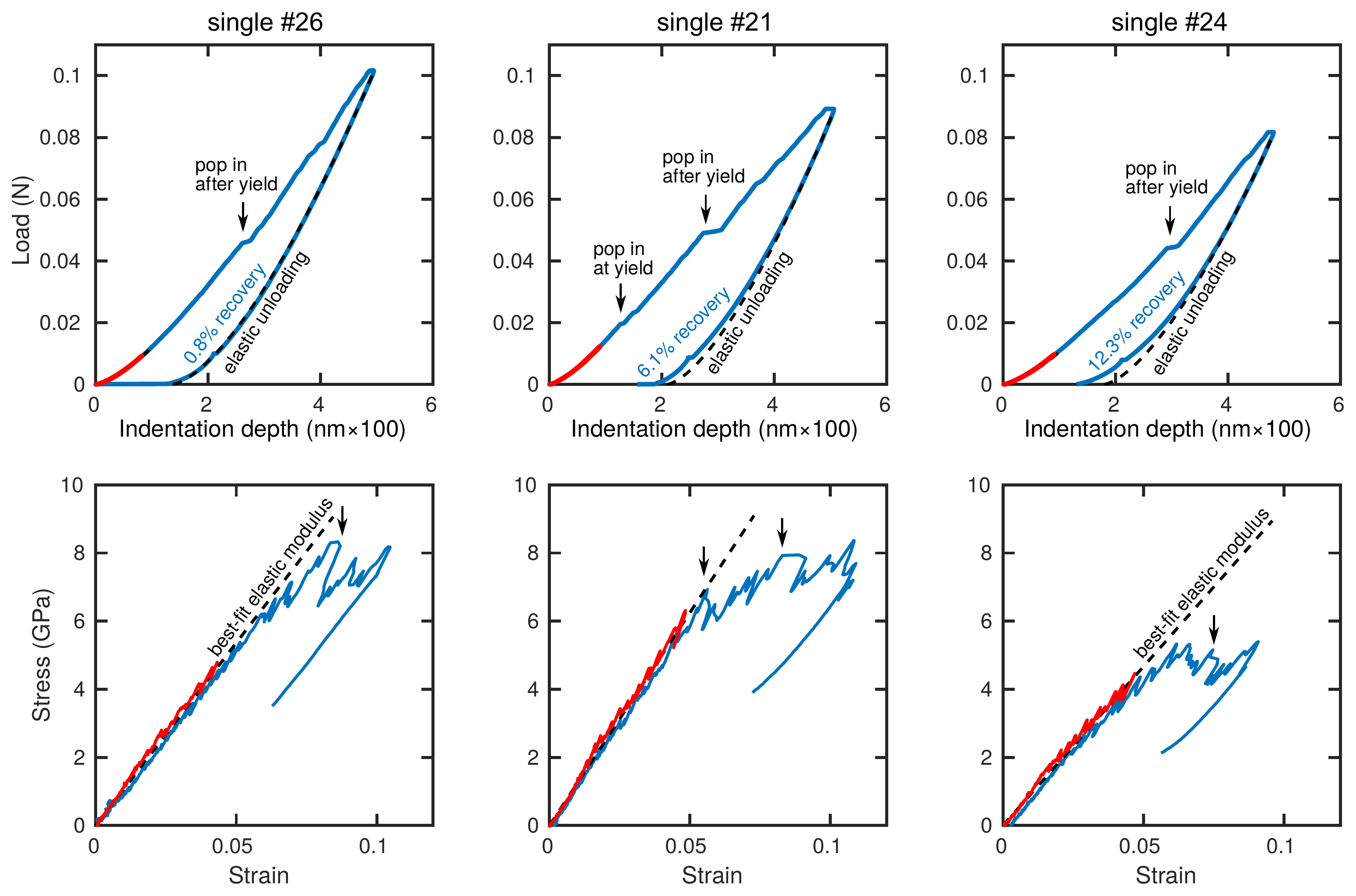}
  \caption{%
    Mechanical data from single indents in first area of interest. Three indent locations are presented with increasing degrees of inelastic strain recovery during unloading. The indentation direction relative to the crystal orientaition is given in Figure 7. Red curves correspond to totally elastic indents to 100 nm depth. Blue curves correspond to indents to 500 nm depth. In the top row, black dashed lines indicate predicted elastic unloading. In the bottom row, the black dashed lines indicate the best-fit elastic modulus during loading. A variety of pop-in events are indicated by black arrows.}
  \label{fig:3}
\end{figure}

Examples of mechanical data from three indents from the arrays of single indents are presented in Figure \ref{fig:3}. Data from initial, shallow indents are presented in red. Load is presented as a function of the total indentation depth (top row of Figure \ref{fig:3}), and shallow indents demonstrate that loading and unloading paths are identical, indicating purely elastic behavior. These load-depth data are presented as stresses and strains in the bottom row of Figure \ref{fig:3}. Red curves are linear in these plots, again indicative of linear elastic behavior, and are parallel to black dashed lines, which indicate the best-fit elastic moduli. Values of the measured elastic moduli are discussed in section \ref{sec:orient}.

Data from the second array of deeper indents are presented in blue. Load and depth data exhibit different loading and unloading paths with residual indentation on the order of 100 nm, indicating appreciable inelastic deformation. Stress-strain curves demonstrate that the loading path departs from linear elasticity at a distinct yield point. Values of the measured yield stresses are discussed in section \ref{sec:orient}. Most loading curves exhibit strain hardening after yield. Load-depth curves also exhibit multiple, near-instantaneous bursts of displacement, often referred to as ``pop-ins''. Several larger pop-ins are indicated with black arrows in Figure \ref{fig:3}. Pop-ins typically only occur after yield, although they do occasionally coincide with yield. In stress-strain curves, pop-ins appear as bursts of strain at constant stress, immediately followed by a stress drop along a path matching the elastic modulus.

A notable feature of these indents is the upward curvature of the load-displacement curves during unloading. Although some upward curvature is expected due to the non-linear nature of Hertzian contacts (Equation \ref{eq:P}), the observed curvature is often more pronounced than expected. As described in Section \ref{sec:process}, we fit an elastic unloading curve to the initial segment of unloading data (see black dashed lines in the top row of Figure \ref{fig:3}). The predicted indentation depth at which the load is zero represents an estimate of the inelastic indentation depth, and the predicted amount of recovered displacement represents the elastic indentation depth, he, at the end of loading. In some cases, the extrapolated unloading curve reasonably matches the unloading data, for which the unloading portion of the stress-strain curves is a straight line with a slope matching the elastic modulus (Figure \ref{fig:3}, indent \#26). However, there are a variety of cases in which more displacement was recovered than expected by the extrapolation, for which the unloading portion of stress-strain curves exhibits upward curvature, indicating a departure from elastic unloading (Figure \ref{fig:3}, indents \#21 and \#24). In this figure, we present the percentage of recovery, which is defined as the amount of unexpected additional recovery normalized by the total inelastic displacement during loading.
  
\subsection{Cyclical indents}

\begin{figure}
  \includegraphics[width=\textwidth]{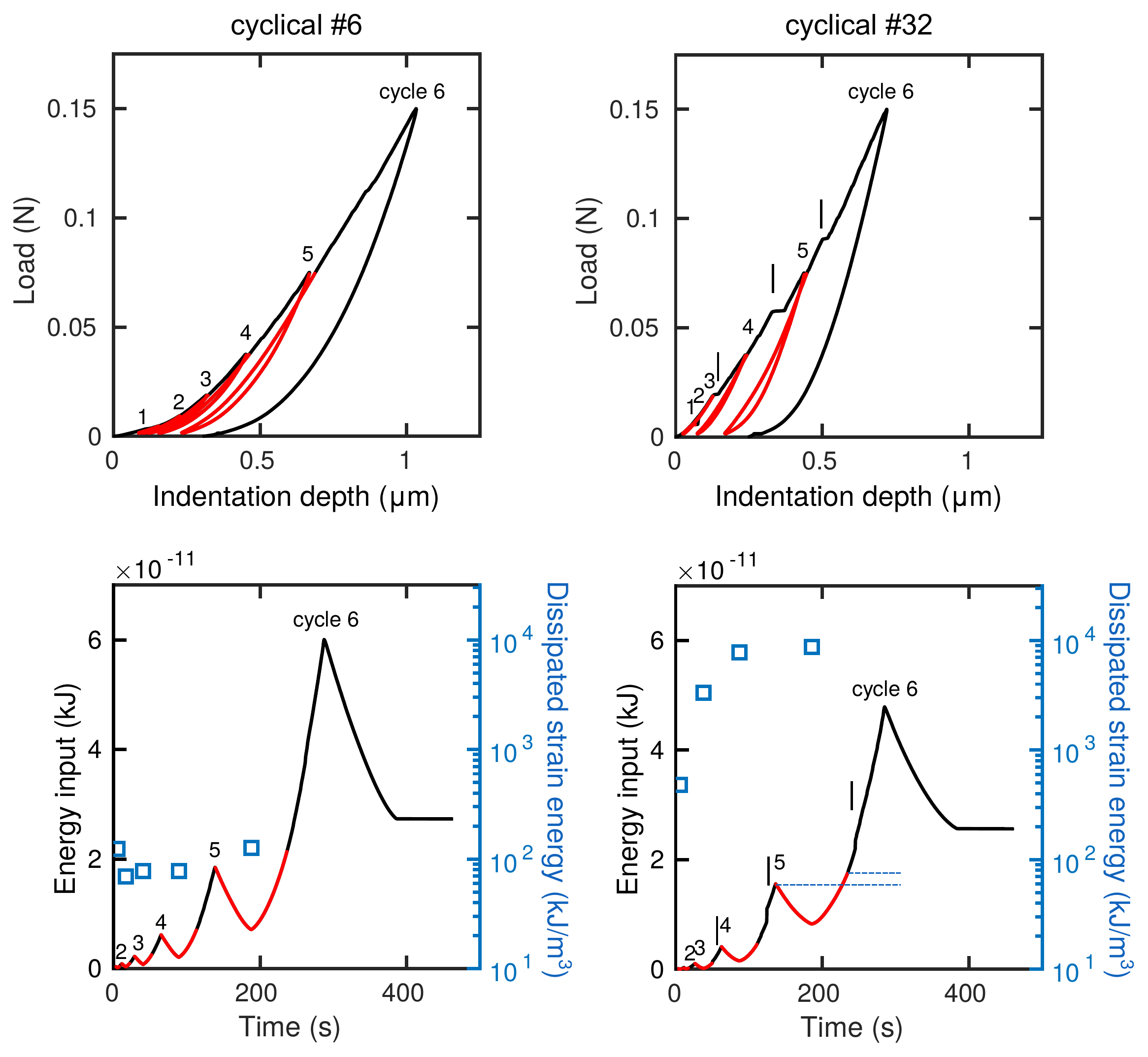}
  \caption{%
    Mechanical data from cyclical indents in second area of interest. The indentation direction relative to the crystal orientatiton is given in Figure 8. Segments highlighted in red indicate unloading and reloading paths that should be identical if the deformation is totally elastic. Two indent locations are presented, one with a small difference between loading and unloading paths (left) and one with a large difference between loading and unloading paths (right). The top row presents load as a function of total indentation depth. The bottom row presents the total energy input as a function of time (black curves). Blue squares indicate the difference between the energy recovered during unloading and the energy input on subsequent reloading. An example of this difference is indicated by the blue dashed lines. Individual cycles are numbered. Black arrows indicate larger pop-ins.}
  \label{fig:4}
\end{figure}

Examples of mechanical data from two cyclical indents are presented in Figure \ref{fig:4}. These data sets exhibit similar characteristics to those observed in single indents. The first two or three cycles exhibit reversible load-displacement paths, indicating purely elastic behavior. Cycles to larger load amplitudes exhibit residual displacements, indicating the onset of inelastic behavior. These higher-amplitude cycles are also often characterized by pop-ins and significant plastic-strain recovery during unloading.
  
We further characterize the mechanical behavior of cyclical indents by analyzing the energy budget throughout the series of loading cycles. Progressive loading cycles input increasingly larger amounts of energy. For smaller amplitude cycles, nearly all the energy is recovered on unloading, again characteristic of elastic deformation. For larger amplitude cycles, only a fraction of the input energy is recovered during unloading. Notably, the unloading path is different from the subsequent reloading path up until the load equals the maximum load of the previous cycle. At this point, the load-displacement curves for unloading and subsequent reloading coincide again. We calculate the difference between the amount of energy that is recovered on unloading and the amount of energy input on the subsequent reloading. The data used for this calculation are presented in red in Figure \ref{fig:4}. We normalize this dissipated strain energy by the volume of deforming material under the indent inferred from the calculated contact radius (blue squares in Figure \ref{fig:4}). For cycles that are near totally elastic, the energy difference is approximately $10^2$~kJ/m$^3$, which represents the smallest magnitudes resolvable by this technique. Progressively larger amplitude cycles result in increasing differences in dissipated strain energy, with maximum observed values near $10^4$~kJ/m$^3$.

\subsection{Mechanical response as a function of crystal orientation}
\label{sec:orient}

\begin{figure}
  \centering
  \includegraphics[width=\textwidth]{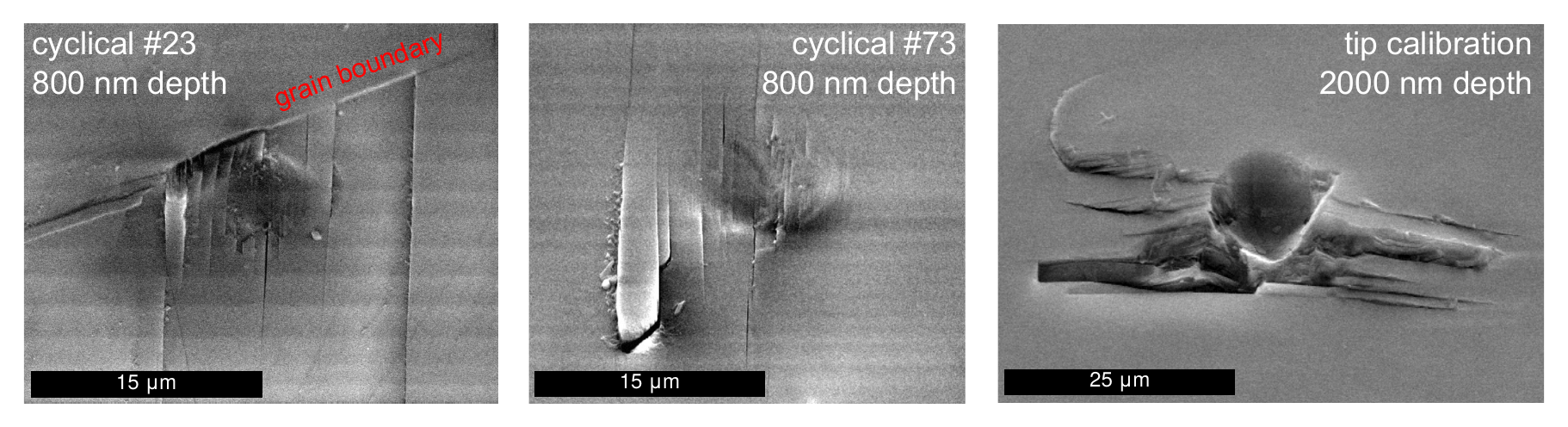}
  \caption{%
    Secondary-electron images of individual indents. The sample is tilted at 70$^\circ$ to electron beam and tilt correction is applied. The maximum depth of indentation is given for each image. Two images are from cyclical loading indents, and their numbers correspond to numbering in Figure \protect\ref{fig:2}. The third is from a calibration indent that was taken to a much greater depth than indents in the two primary regions of interest.}
  \label{fig:5}
\end{figure}

To aid our interpretation of the micromechanical behavior of antigorite, we further consider the mechanical data in the context of the crystallographic orientation at the location of each indent. As an initial assessment, we collected secondary-electron images of several residual indents to examine the surface expression of antigorite deformation, as illustrated in Figure \ref{fig:5}. For imaging, we chose several of the deepest indents, including an indent used for calibrating the indent location, since these indents exhibit the most visible features. In all cases, residual indents are characterized by significant crack formation along the antigorite basal plane. Cracks appear to be primarily mode II with shear offsets visible at the sample surface.
  
\begin{figure}
  \centering
  \includegraphics[width=\textwidth]{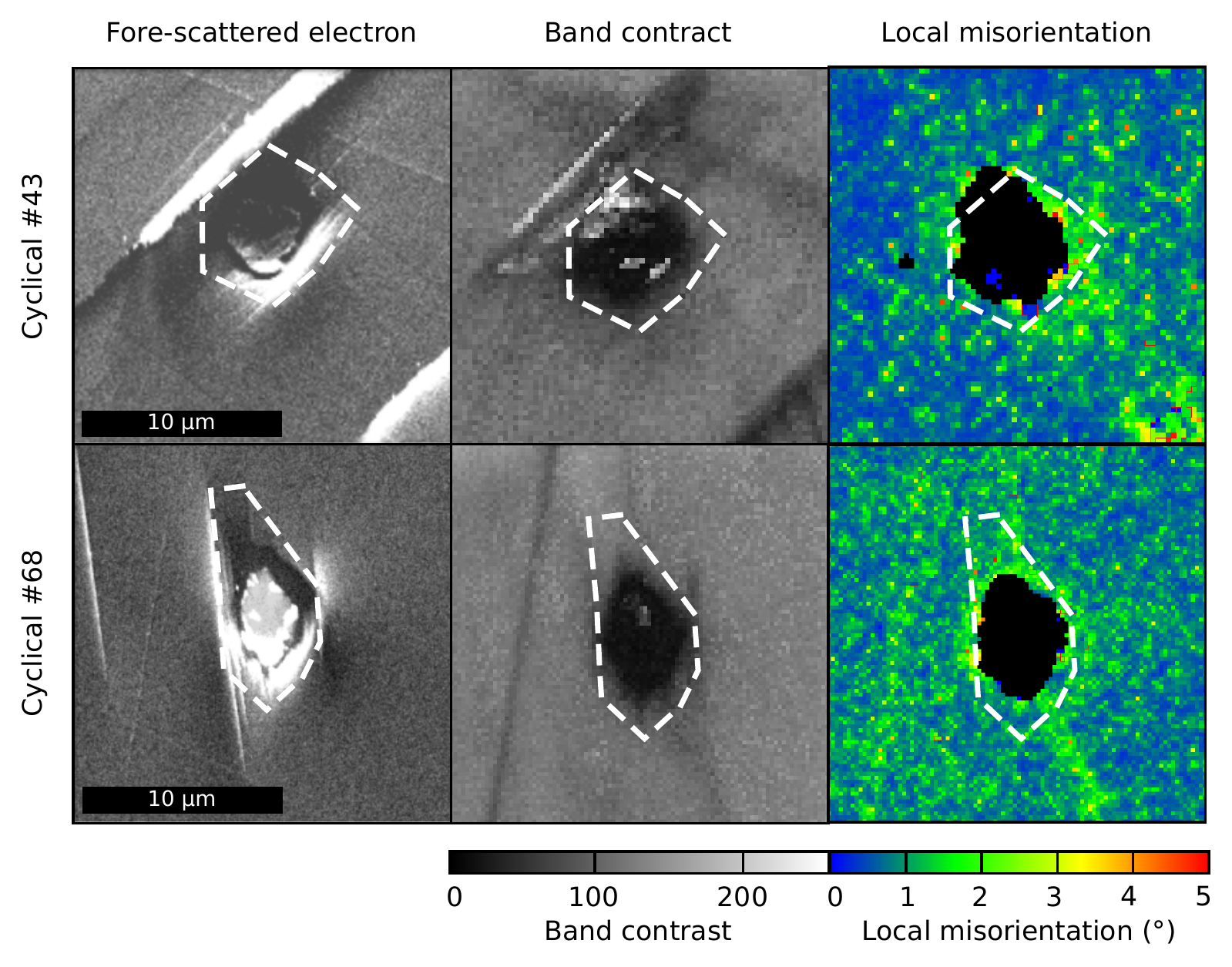}
  \caption{%
    EBSD data local to the residual impressions of two cyclical indents. Images are presented from fore-scatter detectors, band contrast of EBSD patterns, and local misorientation. White dashed lines represent the approximate extent of the residual impression. Local-misorientation maps reveal relatively little signal, with the highest values occurring inside the residual impression and associated with surface damage of the sample.}
  \label{fig:6}
\end{figure}

Additional images of individual indents mapped with EBSD are presented in Figure \ref{fig:6}. Forescatter images again reveal the presence of shear cracks at the surface and parallel to the basal plane. Band-contrast maps reveal that diffraction patterns were degraded within the indent, presumably associated with surface damage due to the indentation process. Some additional degradation is associated with shear cracks and scratches. Importantly, there is little to no degradation surrounding the indent, contrary to the expectation if significant crystal-plastic deformation had occurred. Similarly, there is no measurable distortion of the crystal lattice outside the residual impression, as revealed by the maps of local misorientation. The magnitude of observed misorientations is on the order of the noise level for this type of measurement in antigorite.
  
\begin{figure}
  \centering
  \includegraphics[width=\textwidth]{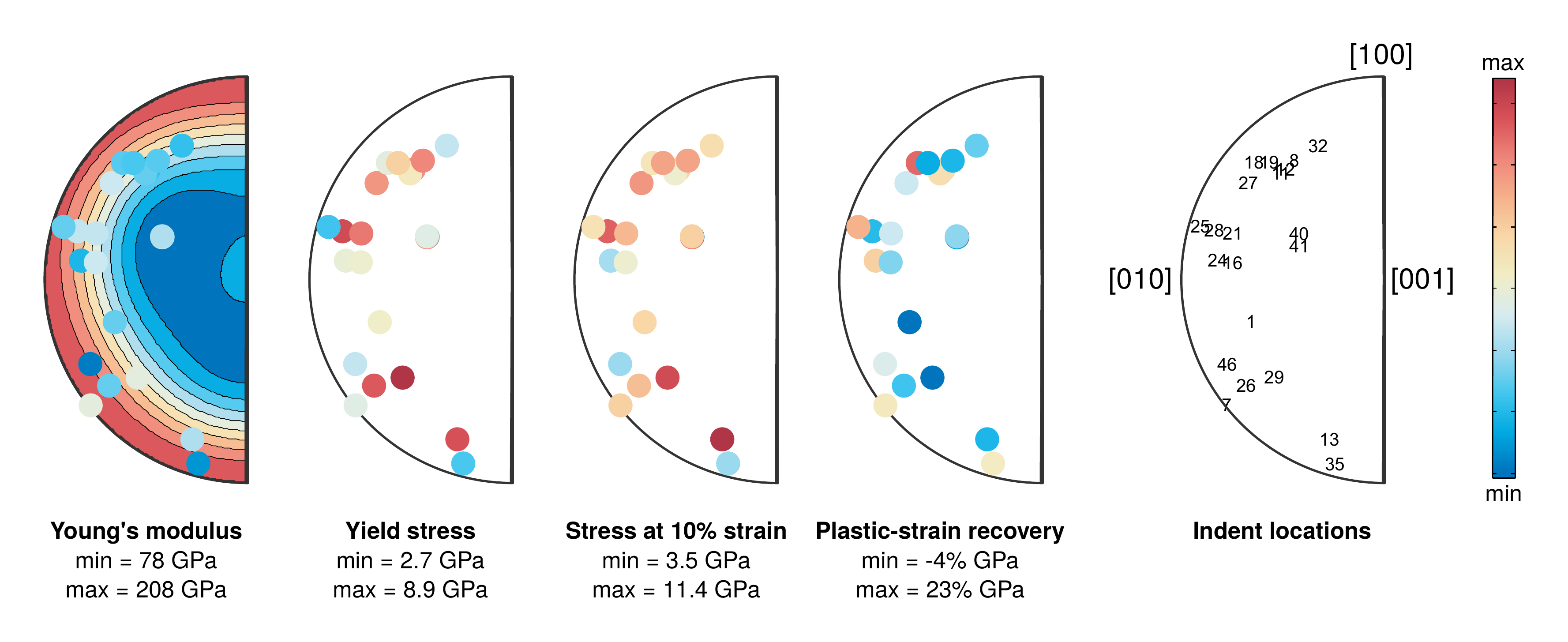}
  \caption{%
    Mechanical behavior of antigorite from single indents as a function of crystal orientation. Indentation directions are plotted in the crystal reference frame using inverse pole figures (IPFs). Data points are colored according to the measured value of modulus, yield stress, stress at 10\% strain, or the magnitude of plastic-strain recovery. Indent locations reference the numbering scheme presented in Figure \protect\ref{fig:2}. The background contouring for Young's modulus is calculated using the elastic constants of \citet{bezacier10}.}
  \label{fig:7}
\end{figure}

\begin{figure}
  \centering
  \includegraphics[width=\textwidth]{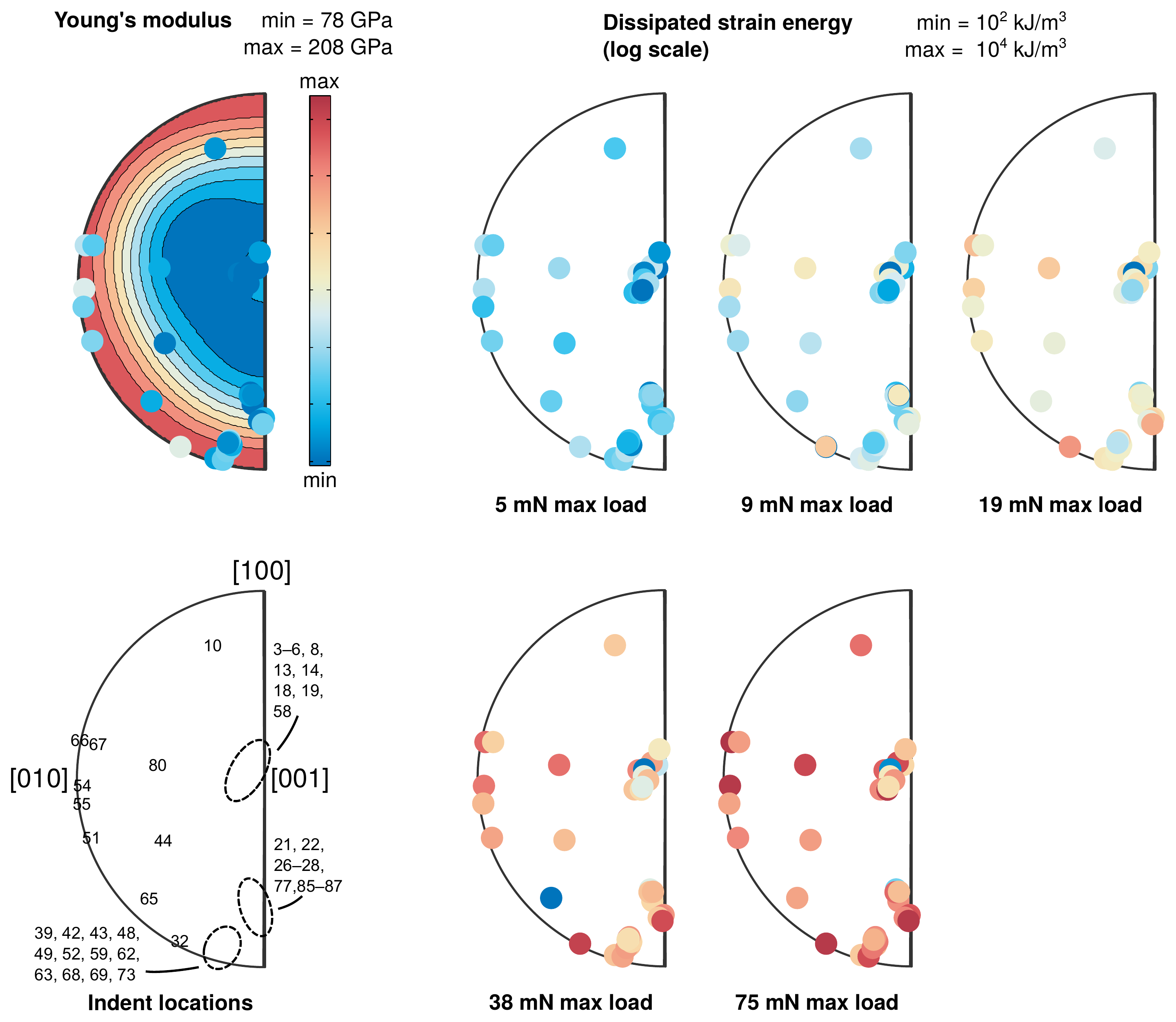}
  \caption{%
    Mechanical behavior of antigorite from cyclical indents as a function of crystal orientation. Indentation directions are plotted in the crystal reference frame using inverse pole figures (IPFs). Data points are colored according to the measured value of modulus and the magnitude of dissipated inelastic strain energy. Magnitudes of dissipated energy are presented on a separate IPF for each loading cycle, and the maximum load in that cycle is noted. Indent locations reference the numbering scheme presented in Figure \protect\ref{fig:2}. The background contouring for Young's modulus is calculated using the elastic constants of \citet{bezacier10}.}
  \label{fig:8}
\end{figure}

We further characterize the role of crystal orientation in the indentation behavior of antigorite by plotting mechanical data in the crystal reference frame. A series of inverse pole figures (IPFs) are presented for single indents (Figure \ref{fig:7}) and cyclical indents (Figure \ref{fig:8}). For single indents, we investigate the measured elastic modulus, the yield stress, the flow stress at 10\% strain, and the magnitude of inelastic-strain recovery. We compare the measured elastic modulus to the Young's modulus measured by \citet{bezacier10} using Brillouin spectroscopy. Although \citet{marquardt15} suggested that \citet{bezacier10} may have mistakenly switched the stiffnesses along the [100] and [010] axes, this difference has little effect on our analysis, and we use the originally report values.  Our measured elastic moduli range from 74 to 132 GPa, and relative to the crystal orientation, these values are generally intermediate to the extremes of the previously published elasticity tensor. This reduced anisotropy in our data is characteristic of spherical indentation \citep{kumamoto17}, which induces a variety of out of plane stresses that result in strains in other crystallographic directions. However, some anisotropy is still evident in the other measured parameters presented in Figure \ref{fig:7}. The lowest yield stresses and lowest stresses at 10\% strain tend to be associated with indents nearly parallel to the basal plane. Furthermore, the highest magnitudes of inelastic-strain recovery also tend to be associated with indents nearly parallel to the basal plane. 
  
Cyclical indents also exhibit a similar dependence of mechanical behavior on crystallographic orientation. Figure 8 presents a series of IPFs with Young's moduli and magnitudes of dissipated strain energy observed in cyclical indents. Similar to single indents, our measured values of Young's modulus are generally intermediate to the extreme values from previously published results for single-crystal antigorite. The exception, however, is the set of indents near parallel to [001], which is an orientation not sampled by single indents. This crystallographic direction is predicted to be the most compliant, and indents in this direction do tend to have the lowest values in our data set, matching published magnitudes. Furthermore, the magnitude of dissipated strain energy tends to be highest for indents parallel to the basal plane, and lowest for indents normal to the basal plane. This pattern is perhaps most distinct for intermediate amplitude cycles (i.e., maximum loads of 19 or 38 mN). At the highest amplitudes, most indents exhibit magnitudes of dissipation near the maximum observed, although indents near perpendicular to the basal plane still exhibit the lowest values.
  
\section{Discussion}
\subsection{Deformation mechanisms during indentation of antigorite}

Our results provide insight on the mechanisms of deformation in antigorite. The key observations include: (1) shear cracks in Figure 5 appear to accommodate deformation at the sample surface; (2) shear cracks are parallel to the basal plane, which is the dominant cleavage plane in antigorite; (3) the lack of crystal distortion surrounding residual indents suggests a paucity of crystal-plastic deformation; and (4) the IPFs in Figure 7 reveal that yield occurs most easily for indents parallel to the basal plane, while it is difficult to initiate yield in indents normal to the basal plane. Taken together, these observations suggest that the basal plane in antigorite is weak and prone to shear microcracking, and that slip along the basal plane is likely the dominant deformation process in our experiments. As demonstrated in Section \ref{sec:model}, deformation by sliding shear cracks along basal planes is also compatible with the inelastic strain recovery observed in single indents (Figure \ref{fig:7}) and  the strain-energy dissipation observed in cyclical indents (Figure \ref{fig:8}).

Slip along basal planes in antigorite can occur in two, possibly nonexclusive, ways: (i) dislocation glide, and (ii) shear fracturing. By contrast with dislocation glide, shear fracturing implies bond breakage,  delamination, and frictional slip along (001) planes, with irreversible loss of cohesion between the slipped portions of the crystal. Our SEM observations indicate that shear cracking did occur during indentation, but are not sufficient to rule out the activity of dislocations entirely.  Previous work investigating the potential for plasticity in antigorite has suggested that dislocation glide would dominantly be on (001) or on conjugate planes that result in apparent slip on (001) \citep[e.g.,][]{amiguet14}, and dislocation interactions are an often cited mechanism for the buildup of backstresses and associated inelastic strain recovery. In addition, even if dislocation glide occurs during unloading, cracks can also form due to stress associated with dislocation interactions \citep{kumamoto17}. Thus, observations 1 and 2 listed above could potentially explained by dislocation activity. However, dislocation activity during indentation tends to result in a halo of geometrically necessary dislocations and associated lattice distortion surrounding residual indents (see Figures 4 and 5 in \citet{wallis19}). In contrast, Figure \ref{fig:6} reveals ano resolvable lattice misorientation or changes in diffraction-band contrast around indents, suggesting there is little to no dislocation accumulation. In addition, dislocation glide in antigorite is also assumed to be predominantly in the [100] direction, and although we observe weak basal planes, we do not observe any directional dependence of the yield stress for indents within the basal plane.

Based on the lack of evidence for dislocation activity, we suggest that our results are most consistent with deformation during unloading being accommodated primarily by sliding on shear cracks. Cracks are often observed around indents on brittle materials, but those are generally related to decompression during unloading. Cracks associated with unloading should be roughly parallel to the surface and normal to the primary tensile stresses. In contrast, the shear cracks we observed are normal to the surface and optimally oriented for shear during loading \citep{swain76}.

The observed pop-ins at or subsequent to yield provide some additional information on the defects available for inelastic deformation. As loading progresses during spherical indentation, not only does the nominal stress increase, but so does the volume of stressed material. Because the stresses at which pop-ins occur are often stochastic in nature, pop-ins are commonly interpreted to reflect the point at which the stress field reaches an available defect source \citep[e.g.,][]{kumamoto17}. Therefore, we suggest that the defects from which shear cracks nucleate in our samples are of low enough density that they are not immediately sampled. The idealized volume of deformation (Figure \ref{fig:1}) at the point of initial pop-in is typically on the order of 1 $\mu$m$^3$ beneath our indents, suggesting that the initial flaw density is approximately 1 $\mu$m$^{-3}$.

\subsection{Elastic moduli, energy dissipation, and recovery during cyclic loading: interpretation based on a sliding crack model}
\label{sec:model}

Cyclical loading experiments provide estimates of the Young's modulus and dissipated strain energy per cycle, both as a function of crystal orientation and the stress amplitude (Figure \ref{fig:8}). To further test our interpretation that inelasticity and anisotropy in antigorite is primarily caused by sliding motion in the basal plane, we develop a simple two-dimensional analytical model that calculates Young's modulus, inelastic strain, and dissipated energy as a function of stress. This ``crack sliding'' model is based on the previously derived formalism of \citet{kachanov82a}, \citet{nemat-nasser88} and \citet{basista98} without crack growth. This formulation is a direct extension of the model of \citet{david12} to triaxial stress and cyclic loading. However, here we consider that all crack faces are initially in contact, an assumption that seems reasonable for sliding along the basal plane in antigorite grains.

The stress conditions during spherical indentation are best described by those of proportional loading, as can be verified by examining the full Hertzian solutions for the stress field inside the loaded body \citep[e.g.,][]{ming16}. Considering the area of primary deformation given in Figure \ref{fig:1}, we find that the average lateral stress 𝜎r is proportional to the average axial stress $\sigma$, i.e., $\sigma_\mathrm{r} = k\sigma$, where the constant $k$ is numerically found to be approximately equal to the Poisson's ratio of the material. Taking $\nu=0.26$ for isotropic antigorite at room temperature
\citep{bezacier10}, $k = 0.25$.

We consider the representative area, $A$, of primary deformation (gray region in Figure \ref{fig:1}) to contain an array of $N$ cracks, each of length $2c$ with its normal oriented at an angle $\phi$ to the applied stress. For simplicity of analysis and availability of analytical solutions, we assume that crack interactions are negligible and that cracks are embedded into an isotropic solid characterized by a single value of the Young's modulus, $E_0$, and of Poisson's ratio, $\nu$.

The essence of the model is that, for crack sliding to be initiated, the applied shear stress must overcome a certain shear strength. Two different cases are considered: (1) ``frictional'' crack sliding, in which the resolved shear stress on a crack must exceed a normal stress-dependent, Coulomb-type frictional resistance and (2) ``cohesive'' crack sliding, in which the resolved shear stress on a crack must exceed a constant yield stress. The cohesion term can arise from several possible physical mechanisms. One possibility is that the cohesion corresponds to the Peierls stress for moving dislocations. However, cohesion resulting from bond breakage across a fracture surface is more consistent with the interpretations presented in Section 4.1. In both frictional and cohesive cases, as sliding proceeds, elastic energy is stored in the material surrounding the cracks. This stored energy leads to the observed strain hardening after yield. During unloading, this stored elastic energy provides a restoring force that promotes backsliding, which is initiated if the sum of the applied shear stress and the elastic restoring force overcomes the sliding shear strength in the reverse direction \citep[e.g.,][]{nemat-nasser88}. The activation of sliding and backsliding at different applied stresses during loading and unloading results in dissipation of strain energy and produces hysteresis in stress-strain curves.

Details of the model and derivations are given in the Appendix. We focus here on two key model outputs: the stress-dependent, effective Young's modulus $E$ and the dissipated energy per unit volume $W$ during unloading from a maximum stress $\sigma^*$ and subsequent reloading to the same stress. For frictional crack sliding (case 1), the effective Young's modulus once crack sliding is initiated is given by
\begin{equation}\label{eq:E1}
  \frac{1}{E} = \frac{1}{E_0}\left[1+\pi\Gamma\sin(2\phi)M_\mathrm{L}\right],
\end{equation}
and the strain energy dissipated per cycle is expressed as
\begin{equation}\label{eq:W1}
  W = \frac{\sigma^{*2}}{2E_0}\left[ \pi\Gamma\sin(2\phi)M_\mathrm{L}\left(1-\frac{M_\mathrm{L}}{M_\mathrm{U}}\right)\right]
\end{equation}
where $\sigma^*$ is the maximum stress, $\Gamma$ is the crack density defined as $\Gamma=Nc^2/A$. $M_\mathrm{L}$ and $M_\mathrm{U}$ are geometrical factors for loading and unloading, respectively, given by
\begin{equation}
  \begin{array}{rl}
    M_\mathrm{L} &= (1-k)\cos\phi\sin\phi -\mu[\cos^2\phi+k\sin^2\phi]\\
    M_\mathrm{U} &= (1-k)\cos\phi\sin\phi +\mu[\cos^2\phi+k\sin^2\phi],
  \end{array}
\end{equation}
where $k$ is the constant of proportional loading given above and $\mu$ is the friction coefficient. For the case of cohesive sliding (case 2), the effective Young's modulus once crack sliding is initiated is given by
\begin{equation}\label{eq:E2}
  \frac{1}{E} = \frac{1}{E_0}\left[1+\pi\Gamma\sin(2\phi)M\right],
\end{equation}
and the energy dissipated per cycle is expressed as
\begin{equation}\label{eq:W2a}
  W = \frac{2\pi\Gamma\sin(2\phi)\tau_\mathrm{y}}{E_0}\left( \sigma^* - \frac{2\tau_\mathrm{y}}{M}\right)
\end{equation}
if there is backsliding ($\sigma^*\geq 2\tau_\mathrm{y}/M$), where $\tau_\mathrm{y}$ is the constant sliding ``yield stress'', and $M$ is a geometrical factor given by
\begin{equation}\label{eq:M}
  M = (1-k)\cos\phi\sin\phi.
\end{equation}

\begin{figure}
  \centering
  \includegraphics{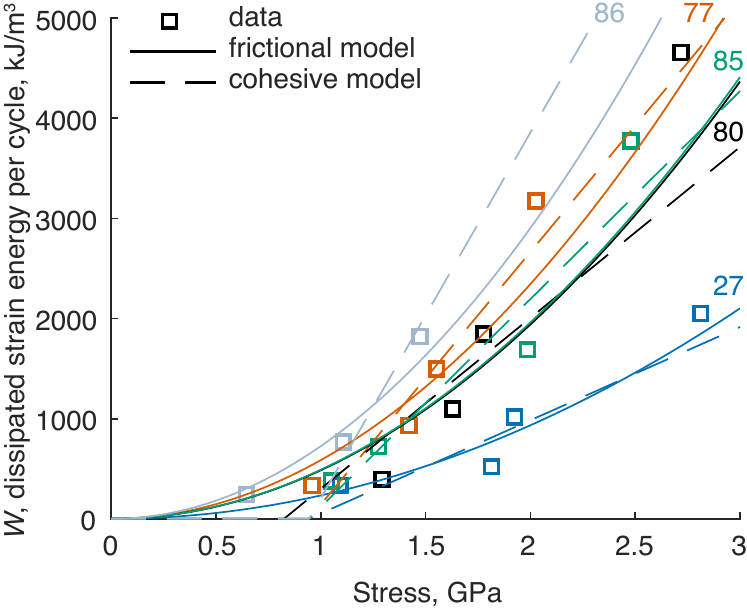}
  \caption{%
    Dissipated strain energy per cycle as a function of the maximum stress per cycle for five cyclical indents. Open squares are observations from indents, and full and dashed curves are best fits to observations using the frictional sliding and the cohesive sliding crack models, respectively. Colors and corresponding numbers indicate the specific indent. Parameters values used in the fitting procedure are discussed in the text.}
  \label{fig:9}
\end{figure}

We invert for model parameters by comparison to our experimental observations of Young's modulus and dissipated strain energy energy for five selected indents (Figure \ref{fig:9}). We assume that all cracks of interest lie in the basal plane (i.e., the crack normal is parallel to [001]). The crack sliding model predicts that sliding is optimal for cracks approximately at an angle of about 60$^\circ$ to the loading direction in the frictional sliding case, and an angle of 45$^\circ$ in the cohesive sliding case. We specifically investigate indents 27, 77, 85, and 86 ($\phi=^\circ$) and indent 80 ($\phi = 52^\circ$), and select cyclical loading data prior to the occurence of any pop-ins events (if observed). For all indents, the Young's modulus of the uncracked solid is taken to be $E_0=97$~GPa, the isotropic Young's modulus for antigorite at room temperature \citep{bezacier10} We only need to invert for two adjustable parameters in each of the models. For the frictional crack sliding model (case 1), the friction coefficient, $\mu$, is imposed to be the same for all indents, and the crack density, $\Gamma$, varies among indent locations. For the cohesive crack sliding model (case 2), the strength, $\tau_\mathrm{y}$, is imposed to be the same for all indents, while the crack density, $\Gamma$, varies among indent locations.

For the two selected crystal orientations, both models are able to jointly fit the modulus deficit relative to the uncracked solid, and the stress-dependent dissipation of energy (Figure \ref{fig:9}). For the frictional crack sliding model, the best fit is obtained with $\mu\approx0.5$ and $\Gamma$ in the range 0.2 to 0.6, depending on indent location. For the cohesive crack model, the best fit is obtained with $\tau_\mathrm{y}\approx0.15$~GPa and $\Gamma$ in the range 0.1 to 0.4. The reasonable fits obtained with both models provides further validation that sliding on shear cracks is a plausible mechanism responsible for inelastic deformation in antigorite. However, a significant difference between the two models is that the dissipated energy $W$ is quadratic in applied stress $\sigma^*$ for the frictional crack sliding model, whereas $W$ is linear in applied stress $\sigma^*$ for the cohesive crack sliding model (at high stress). A qualitative evaluation of Figure \ref{fig:9} suggests the quadratic nature of the frictional sliding crack model is a better representation of the observations, although with the available data, we are currently unable to quantitatively discriminate between the two hypotheses.

\subsection{Comparison to previous work on deformation mechanisms in antigorite}

A variety of deformation mechanisms have been previously proposed to operate in antigorite. Evidence for dislocation-dominated deformation is largely indirect, and most investigators infer that dislocations are the primary means of deformation based on observed CPOs \citep{katayama09,vandemoortele10, padron-navarta12, hirauchi13}. Interpretations of CPOs suggest $(001)$ is the dominant glide plane and $[100]$ is the dominant shear direction. However, microscopic evidence for dislocation activity is inconclusive. Because of the modulated, wave-like nature of the layering in the crystal structure \citep{zussman54}, dislocations may be difficult to generate in the antigorite. The unit cell dimension along $[100]$ is extremely large (35--43~$\AA$), which suggests that line energies for dislocations with $[100]$ Burgers vectors would be high. \citet{otten93} observed stacking defects in these modulations, often referred to as modulation dislocations, but it is unclear whether or not these can be carriers of significant plastic deformation. \citet{amiguet14} observed lattice distortion and kink bands in transmission electron microscope (TEM) images of experimentally deformed antigorite, which they interpreted as a result of dislocation activity. In their analysis, slip was interpreted to appear macroscopically as having occurred on (001), but they suggest that this apparent slip plane was the result of simultaneous slip on $(101)$ and $(10\bar{1})$. \citet{auzende15} also observed structures in TEM images of naturally and experimentally deformed antigorite that they inferred to be built from dislocations. However, based on a range of microstructural observations, they argued that dislocation activity is severely limited in favor of brittle processes. Similarly, recent in situ experiments by \citet{cordier18} did not reveal any dislocation activity during deformation, instead observing delamination and fracture.

Because of the apparent difficulty in activating dislocation slip systems, brittle processes dominate the deformation of antigorite under a wide range of conditions. These brittle processes include cleavage opening near kinks, delamination, and shear microcracking. The link between kinking and cleavage opening in antigorite is supported by the correlation of microstructurally observed kinks and macroscopically observed brittle deformation \citep{nicolas73, chernak10,auzende15, proctor16}. However, recent deformation experiments conducted on aggregates of antigorite at elevated pressure and room temperature reveal that kinking is localized to the damage zone near fractures that formed during brittle failure \citep{david18}. In fact, prior to macroscopic failure, the mechanical behaviour of antigorite aggregates is marked by (1) mostly nondilatant deformation prior to failure in the brittle regime \citep{escartin97,david18}, (2) significant shear dissipation and the absence of volumetric dissipation during cyclic loading at stresses below the brittle failure strength \citep{david18}, and (3) the ubiquitous presence of shear microcracks oriented along the basal (cleavage) planes of antigorite in the ductile (semi-brittle) regime, preferentially orientated at around 45$^\circ$ from the maximum compressive stress \citep{escartin97}. Taken together, these observations support the hypothesis that intragranular shear microcracking, equivalent to shear delamination along the cleavage plane, is a key deformation mechanism in antigorite, at least in the low-temperature regime. Although shear delamination is analogous to, and can be produced by, dislocation glide in the basal plane \citep[e.g.,][]{stroh54}, delamination need not be accommodated by dislocations in antigorite, where intracrystalline slip is likely accommodated by spatially extended defects like shear cracks.

Our results from indentation experiments are consistent with this hypothesis, and the operation of shear cracks can explain both the deformation and microstructural data. Furthermore, since all of our experiments are confined to small regions contained within single crystals, it is clear that shear cracking can occur entirely within grains, suggesting it is possible for shear cracking in macroscopic triaxial experiments to be dominantly intragranular.

\subsection{Implications for antigorite mechanics in Earth}

In nature, for instance within subducting oceanic lithosphere, antigorite deformation occurs at significantly lower strain rates than in laboratory conditions, and in the presence of aqueous fluids. One key limitation of observations from laboratory deformation experiments is that the relatively fast strain rates (and lower temperature) imposed experimentally might limit the mobility of dislocations and thus favor shear cracks, whereas dislocation motion could possibly be dominant at much lower strain rates. The relevance of indentation data and of the shear-cracking mechanism is discussed here by comparison with experimental data obtained on other silicate minerals, and by analyzing the compatibility of shear crack-driven deformation with observed CPOs in naturally deformed antigorite.

Results from indentation tests conducted on quartz \citep[e.g.,][]{masuda00}, olivine \citep[e.g.,][]{evans79,kumamoto17}, and mica \citep[e.g.,][]{basu09} have all demonstrated the activity of one or more dislocation slip systems, even at room temperature, and the rheological behaviour of these minerals determined by indentation tests is consistent with dislocation-dominated deformation mechanisms. Therefore, the self-confining and grain-scale nature of nanoindentation tests typically limits the occurrence of tensile microcracks and instead favors intracrystalline flow mechanisms, even in strong silicate minerals. In contrast, our nanoindents in antigorite do not exhibit detectable dislocations, and our data appear to be well explained by an intragranular shear cracking mechanism. These observations are strong indicators that antigorite does not have any easily activated dislocation glide systems.

While crystal-scale observations of naturally deformed antigorite reveal indirect signs of dislocation activity in the basal plane (stacking disorder, as reported by \citet{auzende15}), they also indicate recrystallisation due to dissolution-precipitation processes, which suggests that antigorite typically does not deform by pure climb-limited dislocation creep, even under natural deformation conditions. Since dissolution-precipitation processes require the presence of aqueous fluids and therefore the existence of some minimal porosity in the rock, evidence for such processes in naturally deformed antigorite supports the hypothesis that some degree of brittle deformation occurs under natural conditions.

Field observations of CPOs in naturally deformed antigorite-rich rocks have commonly been interpreted as the signature of dislocation creep \citep[e.g.,][]{vandemoortele10,padron-navarta12}, with strong implications in terms of the rheology of the material. However, intragranular shear cracking by delamination of the basal planes is not necessarily inconsistent with the development of CPOs, notably by the formation and progressive rotation of antigorite blades by sliding along (001). Such a mechanism for CPO formation has recently been suggested for calcite deformed in the brittle regime \citep{demurtas19}.

At this stage, the mechanism of shear cracking inferred from laboratory deformation experiments at relatively fast strain rates cannot be completely ruled out at geological strain rates, even if CPOs are observed. It remains to be determined how shear cracking interacts with other deformation processes activated at lower strain rates and in the presence of fluids, but we emphasise that intragranular delamination and sliding is likely a significant deformation process under natural conditions.

\section{Conclusions}

We explored the micromechanics of antigorite using instrumented nanoindentation. Spherical indentation was performed on natural samples of antigorite in two separate arrays. In the first, a single loading cycle was performed at each indent location. In the second, multiple loading cycles were performed at each location, with each cycle to a greater maximum load than the previous. Single indents revealed initial elastic loading, a distinct yield point, and strain hardening. During unloading, more strain is recovered than predicted for purely elastic loading. Similarly, cyclical indents recover more strain energy than expected for purely elastic unloading, which was confirmed by examining the difference in energy during unloading and subsequent reloading. This range of mechanical behavior was also observed to be dependent on crystallographic orientation, with lower yield stresses and increased amounts of strain and strain-energy recovery for indents parallel to the antigorite basal plane.

We interpret these mechanical data to reflect sliding on shear cracks along the basal plane. This interpretation is supported by microstructural observations of shear cracks in and surrounding residual indents at the sample surface. We further argue against the activity of dislocations because there is no measurable lattice distortion associated with dislocation accumulation around indents, and there is no apparent preference for sliding direction in the basal plane that might be associated with a Burgers vector. Based on this interpretation, we develop a new microphysical model for an isotropic rock containing an array of sliding cracks that predicts the effective Young's modulus and dissipation of strain energy as functions of the maximum stress. The model is able to successfully explain both the modulus deficit and the dissipated strain energy measured on many indents, with reasonable values of crack density and either friction coefficient (frictional sliding case) or cohesive strength (cohesive sliding case).

\section*{Appendix: calculation of Young's modulus and dissipated strain energy for a rock containing an array of sliding cracks, under proportional loading}

\subsection*{1 Frictional sliding crack}

The ``effective sliding stress'', $\tau_\mathrm{eff}$, driving frictional sliding on loading a crack oriented at angle $\phi$ to the applied stress is the difference between the resolved shear stress on the crack, $\tau$, and the frictional resisting stress, $\tau_\mathrm{f}=\mu\sigma_\mathrm{n}$:
\begin{equation}
  \tau_\mathrm{eff}=\tau-\mu \sigma_\mathrm{n},
\end{equation}
where $\sigma_\mathrm{n}$ is the resolved normal stress and $\mu$ is the friction coefficient. Under proportional loading ($\sigma_\mathrm{r}=k\sigma$, see Section \ref{sec:model}), the projection of the applied stress onto a given crack gives
\begin{equation}
  \tau_\mathrm{eff}=(\sigma-k\sigma)\cos\phi\sin\phi-\mu[\sigma\cos^2\phi+k\sigma\sin^2\phi]=\sigma M_\mathrm{L}
\end{equation}
where $M_\mathrm{L}$ is a function of the crack orientation expressed as
\begin{equation}
  M_\mathrm{L}=(1-k)\cos\phi\sin\phi-\mu[\cos^2\phi+k\sin^2\phi].
\end{equation}
Under the convention that positive stresses are compressive, the condition for sliding on the crack is$\tau_\mathrm{eff}>0$. It is easily demonstrated that an increment of inelastic strain due to an array of sliding cracks is proportional to an increment of the effective sliding stress \citep{nemat-nasser88},
\begin{equation}
  \displaystyle d\epsilon^{\mathrm{i}}=\frac{\pi \Gamma sin(2\phi)}{E_0} d\tau_{\mathrm{eff}} = \pi \Gamma sin(2\phi) M_\mathrm{L} d\sigma
\end{equation}
where $\Gamma$ is the two-dimensional crack density and $E_0$ is the Young's modulus of the solid. The effective Young's modulus, $E$, is obtained by recalling that an increment of total strain is the sum of the elastic strain increment and the inelastic strain increment expressed above. Therefore, for an array of frictional sliding cracks during loading,
\begin{equation}
  \displaystyle \frac{1}{E}=\frac{1}{E_0} \Big[1+\pi \Gamma \sin(2\phi) M_\mathrm{L}\Big].
\end{equation}
During unloading, the effective stress driving backsliding is the difference between the effective sliding stress $\tau^*_\mathrm{eff}$ at the maximum stress (which is the restoring force accumulated during loading), and the joint action of the frictional resisting stress and the resolved applied shear stress, which both act against backsliding \citep{nemat-nasser88}. $\tau_\mathrm{eff}$ is written as
\begin{equation}
  \tau_\mathrm{eff}=\tau_\mathrm{eff}^{\ast}-(\tau+\tau_\mathrm{f})=\tau^{\ast}-\tau_\mathrm{f}^{\ast}-\tau-\tau_\mathrm{f},
\end{equation}
and by projecting the applied stress onto a given crack, $\tau_\mathrm{eff}$ can then be expressed as
\begin{equation}
  \tau_\mathrm{eff}=(\sigma^{\ast}-\sigma)(1-k)\cos\phi\sin\phi-\mu(\sigma^{\ast}+\sigma)(\cos^2\phi+k\sin^2\phi),
\end{equation}
where the condition for backsliding to occur is, similarly to loading, $\tau_\mathrm{eff}>0$. With the same considerations as for the loading case, the effective Young's modulus during unloading is given by
\begin{equation}
E=E_0\quad\text{if} \sigma \geq \sigma^{\ast} (M_\mathrm{L} / M_\mathrm{U})
\end{equation}
if there is no backsliding, or by
\begin{equation}
  \displaystyle \frac{1}{E}=\frac{1}{E_0} \Big[1+\pi \Gamma \sin(2\phi) M_\mathrm{U}\Big]
\end{equation}
if there is backsliding. $M_\mathrm{U}$ is again a function of the crack orientation expressed as
\begin{equation}
  M_\mathrm{U}=(1-k)\cos\phi\sin\phi+\mu[\cos^2\phi+k\sin^2\phi],
\end{equation}
and $\sigma^*(M_\mathrm{L}/M_\mathrm{U})$ is identified as the ``backsliding yield stress'' for a given crack orientation.

By using the relations given above for the effective Young's modulus during loading and unloading, integration of the stress-strain relations allows derivation of the dissipated energy per cycle, $W$, as a function of the maximum stress $\sigma^*$,
\begin{equation}
  \displaystyle W=\frac{\sigma^{\ast 2}}{2E_0}\Big[\pi\Gamma\sin(2\phi)M_\mathrm{L}\Big(1-\frac{M_\mathrm{L}}{M_\mathrm{U}}\Big)\Big].
\end{equation}

\subsection*{2 Cohesive sliding crack}

For the purely cohesive crack case, crack sliding during loading occurs if the resolved shear stress simply exceeds a stress-independent cohesive resistance or ``crack yield stress'', denoted by $\tau_\mathrm{y}$. The ``effective sliding stress'' driving sliding is then simply written as
\begin{equation}
  \tau_\mathrm{eff}=\tau-\tau_\mathrm{y}.
\end{equation}
Following the same conventions as applied for the frictional sliding case above, $\tau_\mathrm{eff}$ under proportional loading is expressed as
\begin{equation}
  \tau_\mathrm{eff}=\sigma (1-k)\cos\phi\sin\phi - \tau_\mathrm{y} = M \sigma - \tau_\mathrm{y},
\end{equation}
where  $M=(1-k)\cos\phi\sin\phi$.

The condition for sliding is again that $\tau_\mathrm{eff}>0$. Using similar considerations as described above for the frictional sliding case, the effective Young's modulus for loading of an array of cohesive sliding cracks is
\begin{equation}
  \displaystyle \frac{1}{E}=\frac{1}{E_0} \Big[1+\pi \Gamma \sin(2\phi) M\Big]\quad\text{for}\quad\sigma > \tau_\mathrm{y} / M
\end{equation}
if there is sliding or
\begin{equation}
  E=E_0
\end{equation}
if there is no sliding, where $\tau_\mathrm{y}/M$ is the ``yield stress'' for a given crack orientation.

During unloading, by analogy to the frictional sliding case, the ``effective backsliding stress'' on a given cohesive crack is
\begin{equation}
   \tau_\mathrm{eff}=\tau_\mathrm{eff}^{\ast}-(\tau+\tau_\mathrm{y})=\tau^{\ast}-\tau-2\tau_\mathrm{y}
\end{equation}
which yields
\begin{equation}
  \tau_\mathrm{eff} = (\sigma^{\ast}-\sigma)(1-k)\cos\phi\sin\phi-2 \tau_\mathrm{y}.
\end{equation}
The condition for backsliding is again that $\tau_\mathrm{eff}>0$. The effective Young's modulus for unloading is then
\begin{equation}
  E=E_0\quad\text{for}\quad\sigma \geq \sigma^{\ast} - 2 \tau_\mathrm{y} / M
\end{equation}
if there is no backsliding or
\begin{equation}
  \displaystyle \frac{1}{E}=\frac{1}{E_0} \Big[1+\pi \Gamma \sin(2\phi) M\Big]
\end{equation}
if there is backsliding, where $\sigma^*-2\tau_\mathrm{y}/M$ is again identified as the ``backsliding yield stress'' for a given crack orientation.

 As described above, the dissipated energy per cycle for the cohesive sliding case is given by
 \begin{equation}
   \displaystyle W=\frac{2\pi \Gamma \sin(2\phi) \tau_{\mathrm{y}}}{E_0}\Big( \sigma^{\ast} -\frac{2 \tau_{\mathrm{y}}}{M}\Big)
 \end{equation}
if there is backsliding during unloading, which corresponds to $\sigma^*\geq2\tau_\mathrm{y}/M$.

\section*{Acknowledgements}

We are grateful to Kathryn Kumamoto and Christopher Thom for useful discussions of nanoindentation. Luiz Morales provided useful insight into EBSD indexing of antigorite. This work was supported by the Natural Environment Research Council through grant NE/M016471/1 to L. Hansen and N. Brantut, and by the European Research Council under the European Union’s Horizon 2020 research and innovation
programme (project RockDEaF, grant agreement \#804685).

\end{document}